\documentclass[aps,prd,twocolumn,showpacs,preprintnumbers,amsmath,floatfix,superscriptaddress]{revtex4-1}
\usepackage{graphicx}
\usepackage{epsfig}
\usepackage{dcolumn}
\usepackage{bbm}
\usepackage[colorlinks,linkcolor=red,anchorcolor=green,citecolor=blue,CJKbookmarks=True]{hyperref}
\usepackage[normalem]{ulem}
\usepackage{amssymb}
\usepackage{tikz,tikz-feynhand}
\usepackage{color}
\begin{document}
\title{Radiative Decay Width of $J/\psi\to \gamma \eta_{(2)}$ from $N_f=2$ Lattice QCD}
\author{Xiangyu Jiang}
\email{jiangxiangyu@ihep.ac.cn}
\affiliation{Institute of High Energy Physics, Chinese Academy of Sciences, Beijing 100049, People's Republic of China}
\affiliation{School of Physics, University of Chinese Academy of Sciences, Beijing 100049, People's Republic of China}

\author{Feiyu Chen}
\affiliation{Institute of High Energy Physics, Chinese Academy of Sciences, Beijing 100049, People's Republic of China}
\affiliation{School of Physics, University of Chinese Academy of Sciences, Beijing 100049, People's Republic of China}

\author{Ying Chen}
\email{cheny@ihep.ac.cn}
\affiliation{Institute of High Energy Physics, Chinese Academy of Sciences, Beijing 100049, People's Republic of China}
\affiliation{School of Physics, University of Chinese Academy of Sciences, Beijing 100049, People's Republic of China}

\author{Ming Gong}
\affiliation{Institute of High Energy Physics, Chinese Academy of Sciences, Beijing 100049, People's Republic of China}
\affiliation{School of Physics, University of Chinese Academy of Sciences, Beijing 100049, People's Republic of China}

\author{Ning Li}
\affiliation{ School of Sciences, Xi'an Technological University, Xi'an 710032, People's Republic of China}

\author{Zhaofeng Liu}
\affiliation{Institute of High Energy Physics, Chinese Academy of Sciences, Beijing 100049, People's Republic of China}
\affiliation{School of Physics, University of Chinese Academy of Sciences, Beijing 100049, People's Republic of China}
\affiliation{Center for High Energy Physics, Peking University, Beijing 100871, People's Republic of China}

\author{Wei Sun}
\affiliation{Institute of High Energy Physics, Chinese Academy of Sciences, Beijing 100049, People's Republic of China}

\author{Renqiang Zhang}
\affiliation{Institute of High Energy Physics, Chinese Academy of Sciences, Beijing 100049, People's Republic of China}
\affiliation{School of Physics, University of Chinese Academy of Sciences, Beijing 100049, People's Republic of China}

\begin{abstract}
    The large radiative production rate for pseudoscalar mesons in the $J/\psi$ radiative decay remains elusive. We present the first lattice QCD calculation of partial decay width of $J/\psi$ radiatively decaying into $\eta_{(2)}$, the $\mathrm{SU(2)}$ flavor singlet pseudoscalar meson, which confirms QCD $\mathrm{U_A(1)}$ anomaly enhancement to the coupling of gluons with flavor singlet pseudoscalar mesons. The lattice simulation is carried out using $N_f=2$ lattice QCD gauge configurations at the pion mass $m_{\pi} \approx 350$ MeV. In particular, the distillation method has been utilized to calculate light quark loops. The results are reported here with the mass $m_{\eta_{(2)}}= 718(8)$ MeV and the decay width $\Gamma(J/\psi\to\gamma \eta_{(2)})=0.385(45)$ keV. By assuming the dominance of $\mathrm{U_A(1)}$ anomaly and flavor singlet-octet mixing angle $\theta=-24.5^\circ$, the production rates for the physical $\eta$ and $\eta'$ in $J/\psi$ radiative decay are predicted to be $1.15(14)\times 10^{-3}$ and $4.49(53)\times 10^{-3}$, respectively, which agree well with the experimental measurement data. Our study manifests the potential of lattice QCD studies on the light hadron production in $J/\psi$ radiative decays.
\end{abstract}
\maketitle

{\it Introduction.---}
The production of light hadrons in the $J/\psi$ radiative decay is mainly through the processes whereby $J/\psi$ annihilates into (at least two) gluons after emitting a photon and the gluons in the final state are hadronized into light hadrons. The abundance of gluons in these processes may favor the production of glueballs, the bound states of gluons, over the conventional light $q\bar{q}$ hadrons based on the naive $\alpha_s$ power counting. Thus, a relatively large branching fraction of a light hadron in the $J/\psi$ radiative decay may indicate that it is a glueball candidate or has a predominant glueball component. Experimentally, pseudoscalar mesons usually have large radiative production rates. For instance, the branching fraction $\mathrm{Br}[J/\psi\to \gamma \eta'(958)]$ is as large as $5.25(7)\times 10^{-3}$~\cite{Zyla:2020zbs}. However, it is known that the $\eta'$ meson is well established as a $q\bar{q}$ meson belonging to the $\textrm{SU(3)}$ flavor nonet made up of the lightest pseudoscalar mesons. Theoretically, the quenched lattice QCD studies~\cite{Morningstar:1997ff,Morningstar:1999rf,Chen:2005mg} predict that the mass of the pseudoscalar glueball is around 2.4–2.6 GeV, which is also confirmed by lattice simulations with dynamical quarks~\cite{Chowdhury:2014mra,Richards:2010ck,Gregory:2012hu,Sun:2017ipk}. These results also disfavor $\eta'$ and other pseudoscalar mesons with masses below $2$ GeV to be glueball candidates. Therefore, their large production rates in the $J/\psi$ radiative decay need to be understood. One possible reason is that the QCD $\mathrm{U_A(1)}$ anomaly nonperturbatively enhances the coupling of gluons with flavor singlet pseudoscalar mesons. One can check this through the dimensionless effective coupling $g_{J/\psi X\gamma}$ which can be extracted by subtracting the kinetic factor from the measured branching fraction of each process $J/\psi\to\gamma X$ (here $X$ refers to a specific pseudoscalar state). It turns out that $g_{J/\psi X\gamma}$ have similar magnitudes for different $X$ and are close to that of the pure gauge pseudoscalar glueball~\cite{Gui:2019dtm}. This observation implies a possible common theoretical mechanism that $\mathrm{U_A(1)}$ anomaly plays a crucial role in the process of $J/\psi$ radiative decaying to pseudoscalars. Although the above discussion may be a reasonable explanation, performing a quantitative derivation of the pseudoscalar production rate is highly desired to confirm this possibility. In this Letter, we present the first lattice QCD calculation of the partial width of the decay process $J/\psi\to \gamma \eta_{(2)}$, where $\eta_{(2)}$ means the isoscalar pseudoscalar meson for $N_f=2$ flavors. Since the underlying gluonic dynamics of $N_f=2$ is very similar to that of $N_f=2+1$, the result of $\eta_{(2)}$ can be easily extended to $\eta$ and $\eta'$ by considering their mixing.

    {\it Formalism.---}
The partial decay width of the process $J/\psi\to \gamma \eta_{(2)}$ is related to the on shell form factor $M(Q^2=0)$ as
\begin{equation}\label{eq:width}
    \Gamma(J/\psi\to \gamma \eta_{(2)})=\frac{4\alpha}{27}|\vec{q}_\gamma|^3 M^2(0),
\end{equation}
where the electric charge of charm quark $Q_c=2e/3$ has been incorporated, $\alpha=1/134$ is the fine structure constant at the charm quark mass scale, and $|\vec{q}_\gamma|=(m_{J/\psi}^2-m_{\eta_{(2)}}^2)/(2m_{J/\psi})$ is the spatial momentum of the final state photon. The form factor $M(Q^2)$ is defined through the electromagnetic multipole decomposition~\cite{Dudek:2006ej} of the transition matrix element in this process, namely,
\begin{eqnarray}\label{eq:matrix}
    && \langle \eta_{(2)}(p_{\eta_{(2)}})|j_\mathrm{em}^\mu(0)|J/\psi(p_{J/\psi}, \lambda)\rangle \nonumber\\
    &=& M(Q^2) \epsilon^{\mu\nu\rho\sigma}p_{J/\psi,\nu}p_{\eta_{(2)},\rho}\epsilon_\sigma(p_{J/\psi},\lambda),
\end{eqnarray}
where $\epsilon_\sigma(p_{J/\psi},\lambda)$ is the polarization vector of $J/\psi$ and $j_\mathrm{em}^\mu=\bar{c}\gamma^\mu c$ is the electromagnetic current of the charm quark. Here we only consider the initial state radiation and ignore photon emissions from sea quarks and the final state. Theoretically, this matrix element can be extracted from the following three-point function
\begin{equation}\label{eq:three-point}
    \Gamma_{\mu i}^{(3)} (\vec{q};t,t')=\sum\limits_{\vec{y}} e^{i\vec{q}\cdot\vec{y}}\langle \mathcal{O}_{\eta_{(2)}}(\vec{p'},t) j_\mathrm{em}^\mu(\vec{y},t')\mathcal{O}_{J/\psi}^\dagger(\vec{p},0) \rangle
\end{equation}
with $\vec{q}=\vec{p'}-\vec{p}$, where $\mathcal{O}_{\eta_{(2)}}(\vec{p},t)$ and $\mathcal{O}_{J/\psi}(\vec{p},t)$ are the interpolating field operators for the ${\eta_{(2)}}$ state and the $J/\psi$ state with spatial momentum $\vec{p}$, respectively. Therefore, the major task is the calculation of $\Gamma_{\mu i}^{(3)} (\vec{q};t,t')$ which can be done directly in the lattice formalism. Since the decay process occurs in the transition from charm quarks to light quarks which is mediated by gluons, quark annihilation diagrams are necessarily involved in the calculation by using the distillation method~\cite{Peardon:2009gh}.

\begin{table}[t]
    \renewcommand\arraystretch{1.5}
    \caption{Parameters of the gauge ensemble.}
    \label{tab:config}
    \begin{ruledtabular}
        \begin{tabular}{lllllc}
            $L_s^3\times T$   & $\beta$ & $a_t^{-1}$ (GeV) & $\xi$      & $m_\pi$ (MeV) & $N_\mathrm{cfg}$ \\\hline
            $16^3 \times 128$ & 2.0     & $6.894(51)$      & $\sim 5.3$ & $348.5(1.0)$  & $6991$           \\
        \end{tabular}
    \end{ruledtabular}
\end{table}

{\it Numerical details.---}We have generated gauge configurations with $L_s^3\times T=16^3\times 128$ anisotropic lattice by using the tadpole improved Symanzik's gauge action~\cite{Morningstar:1997ff,Chen:2005mg} and the tadpole improved clover fermion action for degenerate $u,d$ quarks~\cite{Edwards:2008ja,Sun:2017ipk}. The renormalized anisotropy parameter and the temporal lattice spacing are determined to be $\xi=a_s/a_t=5.3$ and $a_t^{-1}=6.894(51)$ GeV. Therefore, the spatial lattice spacing is $a_s=0.152(1)$ fm~\cite{Jiang:2022ffl}. Pion mass $m_\pi=348.5(1.0)$ MeV is related with the parameter of bare $u,d$ quark mass. The value $m_\pi L_s a_s\approx 3.9$ warrants that the finite volume effects on this lattice setup are not important. We use 6991 configurations to guarantee the good signals of the correlation functions which involve disconnected quark diagrams. For the valence charm quark, we adopt the clover fermion action in Ref.~\cite{CLQCD:2009nvn} with the charm quark mass parameter being tuned to give $(m_{\eta_c}+3m_{J/\psi})/4=3069$ MeV. With this action, the masses of $\eta_c$ and $J/\psi$ are derived precisely to be $m_{\eta_c}=2.9750(3)$ GeV and $m_{J/\psi}=3.0988(4)$ GeV. The parameters of our gauge ensemble are listed in Table~\ref{tab:config}. On each gauge configuration, we generate the perambulators of $u,d$ quarks in the Laplacian Heaviside (LH) subspace spanned by the $N=70$ eigenvectors with the lowest eigenvalues, such that the annihilation diagrams of light quarks can be calculated conveniently~\cite{Peardon:2009gh}.

\begin{figure}[t!]
    \includegraphics[width=0.9\linewidth]{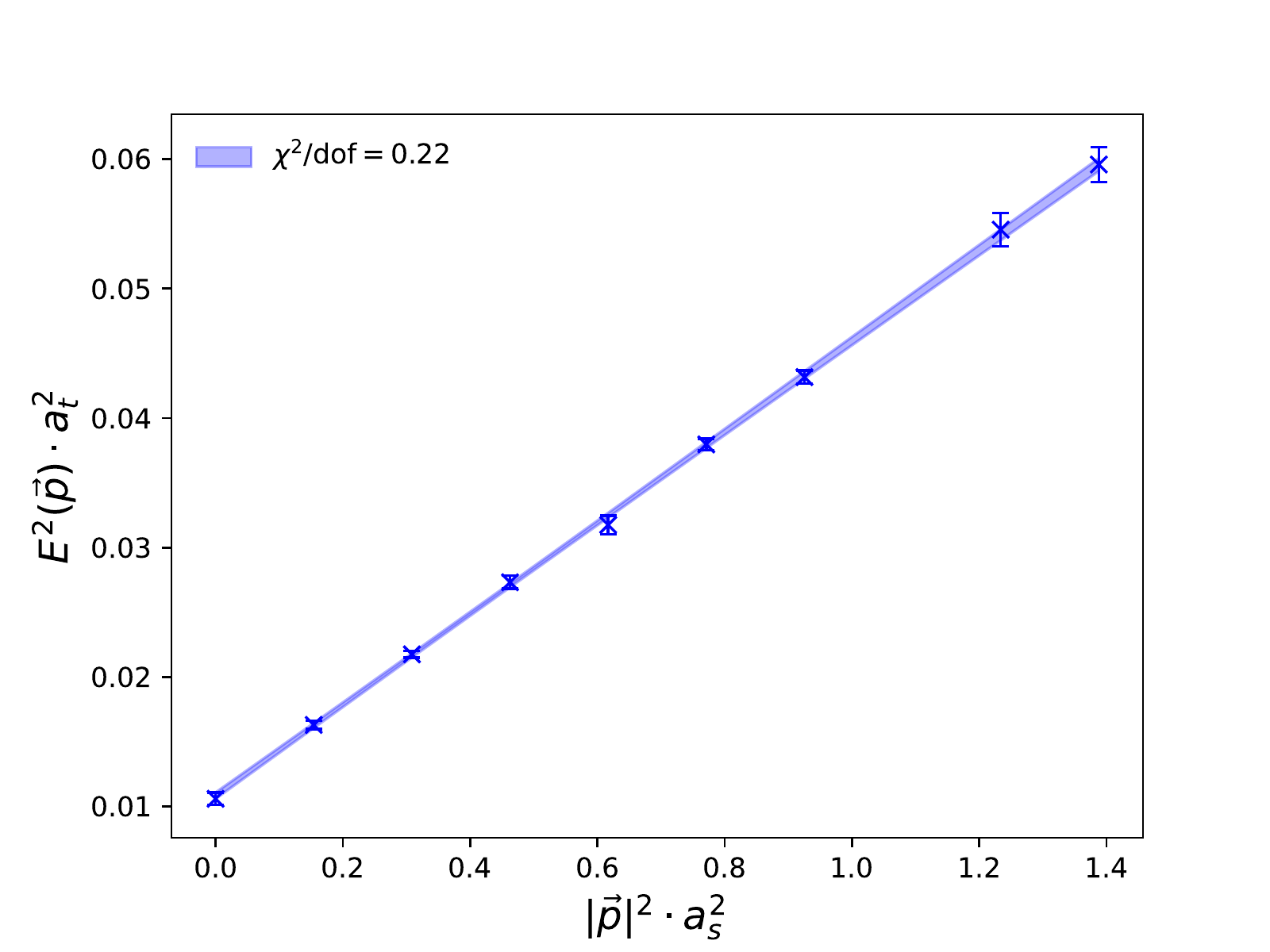}
    \caption{\label{fig:dispersion}The dispersion relation for $\eta_{(2)}$. The data points show the numerical results, and the band exhibits the error of the fitting. The continuum dispersion relation $E^2(\vec{p})a_t^2=m_{\eta_{(2)}}^2a_t^2+\frac{1}{\xi^2}|\vec{p}|^2a_s^2$ is applied, and the fitted parameters are $\xi=5.336(36)$ and $m_{\eta_{(2)}}=718(8)$ MeV.}
\end{figure}

The operator set of the isoscalar ${\eta_{(2)}}$ includes various types of operators. The explicit form is $\mathcal{O}_{\eta_{(2)}}=\frac{i}{\sqrt{2}}\left(\bar{u}\Gamma u+\bar{d}\Gamma d \right)$, where $u$ and $d$ are LH smeared quark fields~\cite{Peardon:2009gh}, and $\Gamma$ refers to $\gamma_4\gamma_5$, $\gamma_4\gamma_5\gamma_i\nabla_i$ or $\gamma_4\gamma_i\mathbb{B}_i$. Here $\nabla=\overrightarrow{\nabla}-\overleftarrow{\nabla}$ with $\overrightarrow{\nabla}$($\overleftarrow{\nabla}$) denoting the gauge covariant derivative acting on quark fields from the right(left) side, and $\mathbb{B}_i$ is the antisymmetric combination of $\nabla_i$. The corresponding operators for a moving $\eta_{(2)}$ with spatial momentum $\vec{p}=\frac{2\pi}{L a_s}\vec{n}$ are obtained through the Fourier transformation. Thus for each $\vec{p}$, we obtain the optimized operator $\mathcal{O}_{\eta_{(2)}}(\vec{p},t)$ which couples most to the lowest state $\eta_{(2)}$ by solving the generalized eigenvalue problem to the correlation matrix of this operator set. In this Letter, the momentum mode $\vec{n}$ runs from $\vec{n}=(0,0,0)$ up to $\vec{n}=(1,2,2)$ to guarantee that the region with $Q^2\sim 0$ can be reached. Although the Fourier transformed operators also couple to a moving meson state with different $J^{PC}$ quantum numbers from $0^{-+}$~\cite{Thomas:2011rh}, it does not matter in the present case since the lowest state for each momentum mode must be $\eta_{(2)}$. This is justified by the correct dispersion relation of $\eta_{(2)}$ shown in Fig.~\ref{fig:dispersion}, where $E(\vec{p})$ are the energies of the ground states contributing to the correlation functions $\Gamma_{\eta_{(2)}\eta_{(2)}}^{(2)}(\vec{p},t)=\langle \mathcal{O}_{\eta_{(2)}}(\vec{p},t)\mathcal{O}_{\eta_{(2)}}^\dagger(\vec{p},0)\rangle$, and the straight line
illustrates the fit using the continuum dispersion relation $E^2(\vec{p})a_t^2=m_{\eta_{(2)}}^2 a_t^2 +\frac{1}{\xi^2}|\vec{p}|^2 a_s^2$ with the best-fit parameters $m_{\eta_{(2)}}=718(8)$ MeV and $\xi=5.336(36)$. 

We use the continuum current form $j_\mathrm{em}^\mu(x)=\bar{c}(x)\gamma^\mu c(x)$ for the electromagnetic current of charm quarks, which is not conserved on the lattice and should be renormalized. We adopt the strategy used in Refs.~\cite{Dudek:2006ej,Yang:2012mya} to determine the renormalization factor $Z_V$. By calculating the relevant electromagnetic form factors of $\eta_c$, we obtain $Z_V^{t}=1.165(3)$ for the temporal component of $j_\mathrm{em}^\mu(x)$ and $Z_V^s=1.118(4)$ for its spatial components.
In this Letter, only $Z_V^s$ is involved and is incorporated into the following expressions implicitly.
\begin{figure}[t!]
    \centering
    \begin{eqnarray}
        \begin{tikzpicture}[> = Stealth, baseline = 0 cm]
            \begin{feynhand}
                \definecolor{gray}{RGB}{170,170,170}
                \definecolor{blue}{RGB}{0,0,255}
                \definecolor{red}{RGB}{255,0,0}
                \vertex[particle] (c) at (0.8,0.5) {};
                \vertex[particle] (cbar) at (0.8,-0.5) {};
                \vertex[dot] (j) at (3,0.5) {};
                \vertex[dot] (g1) at (3,0);
                \vertex[dot] (g2) at (3,-0.5);
                \vertex[dot] (g1_) at (4.2,0);
                \vertex[dot] (g2_) at (4.2,-0.5);
                \vertex[particle] (gamma) at (6,0.8) {$\gamma(-\vec{q},\epsilon_\mu^*)$};
                \vertex[particle] (q) at (5.5,0) {};
                \vertex[particle] (qbar) at (5.5,-0.5) {};
                \graph{
                (c) --[fer] (j)
                --[plain] (g1)
                --[plain] (g2)
                --[fer] (cbar)
                };
                \graph{
                (j) --[pho] (gamma)
                };
                \graph{
                (g1) --[glu,gray] (g1_)
                };
                \graph{
                (g2) --[glu,gray] (g2_)
                };
                \graph{
                (q) --[antfer] (g1_)
                --[plain] (g2_)
                --[antfer] (qbar)
                };
                \filldraw[fill=gray,draw=gray] (0.92,0) ellipse [x radius=0.1, y radius=0.5];
                \filldraw[fill=gray,draw=gray] (5.38,-0.25) ellipse [x radius=0.1, y radius=0.25];
                \vertex[] () at (0,0) {$J/\psi(\vec{0},\epsilon_i)$};
                \vertex[] () at (6, -0.25) {$\eta_{(2)}(\vec{q})$};
                \vertex[blue] () at (0.8,0.8) {$\mathcal{O}_{J/\psi,i}^{\dagger}(\vec{0},0)$};
                \vertex[blue] () at (3,0.8) {$j_\mathrm{em}^\mu(\vec{y},t')$};
                \vertex[red] () at (2,-0.8) {$G_{\mu i}(\vec{q},t,t')$};
                \vertex[blue] () at (5.4,-0.8) {$\mathcal{O}_{\eta_{(2)}}(\vec{q},t)$};
                \vertex[gray] () at (3.6,-0.19) {.};
                \vertex[gray] () at (3.6,-0.26) {.};
                \vertex[gray] () at (3.6,-0.33) {.};
            \end{feynhand}
        \end{tikzpicture} \nonumber
    \end{eqnarray}
    \caption{\label{fig:quark-diagram}The schematic diagram for the process $J/\psi\to \gamma \eta_{(2)}$.}
\end{figure}
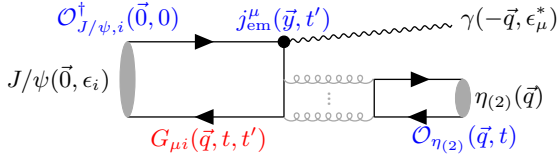

For $J/\psi$ in its rest frame, we use the conventional quark bilinear operator $\mathcal{O}_{J/\psi,i}(\vec{0},t)=\sum\limits_{\vec{z}}\bar{c}(\vec{z},t)\gamma_i c(\vec{z},t)$ in the three-point function $\Gamma_{\mu i}^{(3)} (\vec{q};t,t')$. Since charm quarks and light quarks are contracted separately, for the kinetic configuration where $J/\psi$ is at rest and $\eta_{(2)}$ moves with momentum $\vec{q}$, we re-express $\Gamma_{\mu i}^{(3)} (\vec{q};t,t')$ as
\begin{equation}\label{eq:three-point}
    \Gamma_{\mu i}^{(3)} (\vec{q};t,t')=\frac{1}{T}\sum\limits_{\tau=1}^{T}\langle \mathcal{O}_{\eta_{(2)}}(\vec{q},t+\tau) G_{\mu i}(\vec{q};t'+\tau, \tau) \rangle
\end{equation}
with the block $G_{\mu i}(\vec{q};t'+\tau,\tau)$ defined by
\begin{equation}\label{eq:charm-loop}
    G_{\mu i}(\vec{q};t'+\tau,\tau)=\sum\limits_{\vec{y}} e^{i\vec{q}\cdot\vec{y}}j_\mathrm{em}^\mu(\vec{y},t'+\tau)\mathcal{O}_{J/\psi,i}^\dagger(\vec{0},\tau),
\end{equation}
where we average over all the source time slices $\tau\in [1,T]$ to increase the statistics. The schematic quark diagram of $\Gamma_{\mu i}^{(3)}$ after Wick contraction is shown in Fig.~\ref{fig:quark-diagram}, where the left loop of quark lines is given by the factor $G_{\mu i}$ in Eq.~(\ref{eq:charm-loop}) and the right part is a light quark loop from the self-contraction of $\mathcal{O}_{\eta_{(2)}}$. On each configuration, the two parts are evaluated independently. We remark that the light quark loops can be conveniently calculated through the perambulators of $u,d$ quarks. In order to compute the $G_{\mu i}$ part, which is similar to the calculation of a two-point correlation function for $J/\psi$, we use a wall source to calculate the propagator of charm quark $\sum\limits_{\vec{y}} S_c(\vec{x},t;\vec{y},\tau)$ with $\tau$ running over all the time slices. Thus the charm quark loop in Fig.~\ref{fig:quark-diagram} can be approximated as
\begin{eqnarray}
    &&\sum\limits_{\vec{z}}\mathrm{Tr}\left[\gamma^\mu S_c(\vec{y},t;\vec{z},\tau)\gamma_i\gamma_5 S_c^\dagger(\vec{y},t;\vec{z},\tau)\gamma_5\right]\nonumber\\
    &\to& \sum\limits_{\vec{z},\vec{z'}}\mathrm{Tr}\left[\gamma^\mu S_c(\vec{y},t;\vec{z},\tau)\gamma_i\gamma_5 S_c^\dagger(\vec{y},t;\vec{z'},\tau)\gamma_5\right],
\end{eqnarray}
where the additional terms in the second line are not gauge invariant and will be canceled out after averaging over gauge configurations with enough statistics. It should be emphasized that in the calculation of $G_{\mu i}$, the source operator $\mathcal{O}_{J/\psi,i}(\vec{p},t)$ should have a definite momentum projection (we use $\vec{p}=0$ in this Letter); otherwise, one cannot get available signals for the three-point correlation function $\Gamma_{\mu i}^{(3)}$.

\begin{table*}[t]
    \renewcommand\arraystretch{1.5}
    \caption{The values for $E_{\eta_{(2)}}(\vec{p})$, $Q^2$, and $M(Q^2)$ at different momentum mode $\vec{n}$. All the values are converted into the physical unit using $a_t^{-1}\approx 6.894$ GeV. }
    \label{tab:result}
    \begin{ruledtabular}
        \begin{tabular}{lccccccccc}
            $\vec{n}(\vec{p})$          & (0,0,1)    & (0,1,1)    & (1,1,1)    & (0,0,2)    & (0,1,2)    & (1,1,2)     & (0,2,2)    & (1,2,2)     \\\hline
            $E_{\eta_{(2)}}(\vec{p})$ (GeV) & 0.8801(87) & 1.0167(61) & 1.139(11)  & 1.228(14)  & 1.3434(80) & 1.4324(83)  & 1.610(19)  & 1.683(19)   \\
            $Q^2$ (GeV$^2$)             & $-$4.668(39) & $-$3.819(25) & $-$3.062(42) & $-$2.460(52) & $-$1.782(28) & $-$1.216(28)  & $-$0.132(56) & 0.340(54)   \\
            $M(Q^2)$ (GeV$^{-1}$)       & 0.0490(57) & 0.0291(20) & 0.0222(15) & 0.0187(19) & 0.0161(10) & 0.01301(66) & 0.0123(13) & 0.00973(90) \\
        \end{tabular}
    \end{ruledtabular}
\end{table*}

\begin{figure}[t!]
    \includegraphics[width=0.9\linewidth]{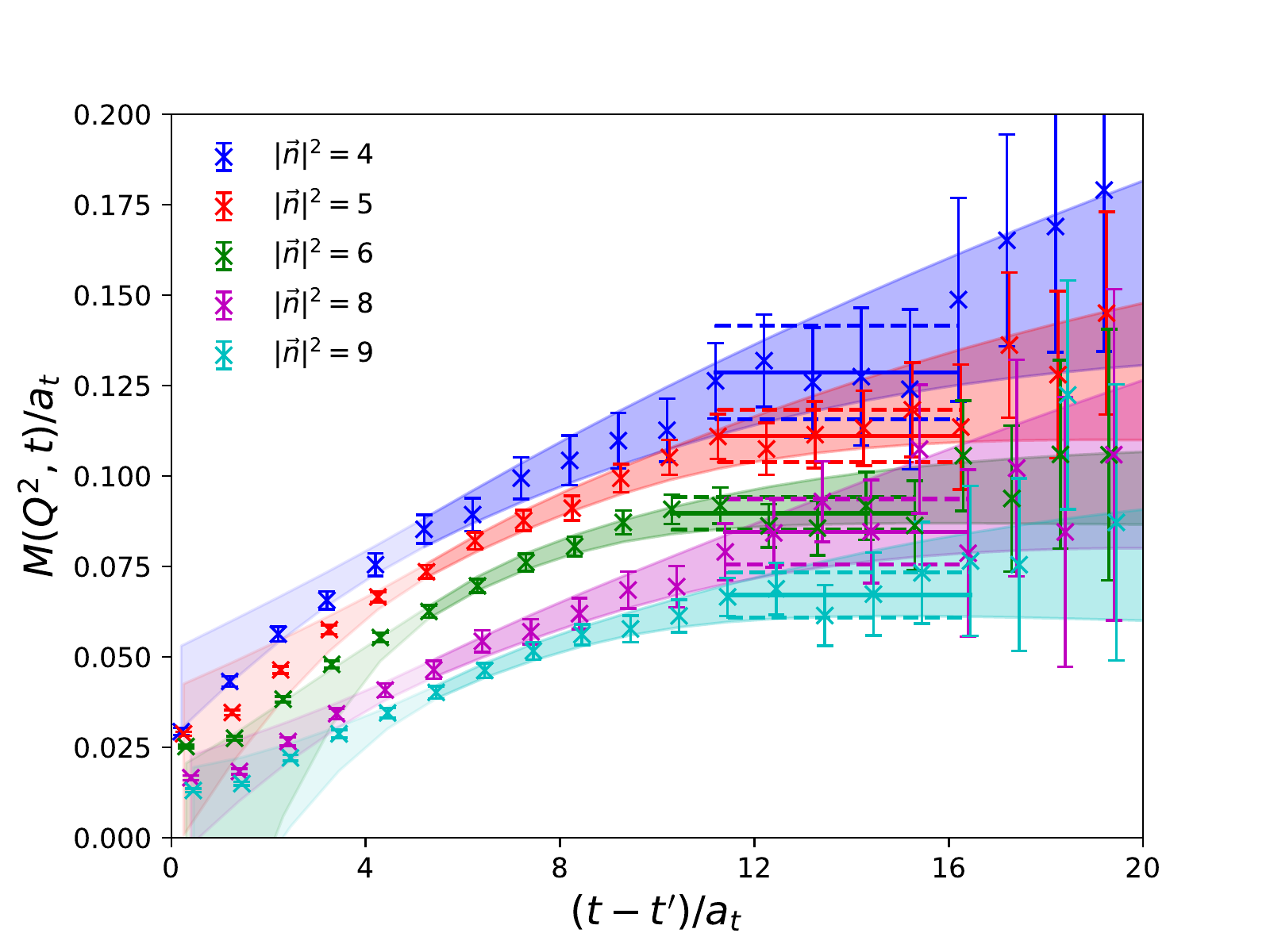}
    \caption{\label{fig:formfactor-t}Form factors $M(Q^2,t)$ versus $t-t'$ with $t'=40$. Different colors indicate different $|\vec{n}|^2$, which lead to some near-zero $Q^2$. The horizontal solid lines along with dashed lines illustrate the fitted values and errors of $M(Q^2)$ with constants as fitting formulas, and the fitting ranges are shown as ranges of these lines. The shaded bands are fitting results to the data points by using $M(Q^2,t)=M(Q^2)+c(Q^2)e^{-\delta E(t-t')}$ as fitting formulas. All the errors are obtained by jackknife resampling.}
\end{figure}

When $t\gg t'$ and $t'\gg 0$, $\Gamma_{\mu i}^{(3)}(\vec{q};t,t')$ can be expressed as
\begin{eqnarray}\label{eq:spectral}
    \Gamma_{\mu i}^{(3)} (\vec{q};t,t')&\approx& \frac{Z_{\eta_{(2)}}(\vec{q})Z_{J/\psi}^{*}}{4V E_{\eta_{(2)}}(\vec{p})m_{J/\psi}} e^{-E_{\eta_{(2)}}(\vec{p})(t-t')}e^{-m_{J/\psi}t'}\nonumber\\
    &\times& \sum\limits_{\lambda}\langle {\eta_{(2)}}(\vec{q})|j_\mathrm{em}^\mu|J/\psi(\vec{0},\lambda)\rangle\epsilon_i^*(\vec{0},\lambda),
\end{eqnarray}
where $V$ is the spatial volume, $Z_{\eta_{(2)}}(\vec{q})=\langle \Omega|\mathcal{O}_{\eta_{(2)}}(\vec{q})|{\eta_{(2)}}(\vec{q})\rangle$, and $Z_{J/\psi}\epsilon_i(\vec{0},\lambda)=\langle \Omega|\mathcal{O}_{J/\psi,i}(\vec{0})|J/\psi(\vec{0},\lambda)\rangle$. The dependence of $Z_{\eta_{(2)}}$ on $\vec{q}$ is due to the LH smeared operator $\mathcal{O}_{\eta_{(2)}}$~\cite{Bali:2016lva}. The two-point correlation functions can be expressed as
\begin{eqnarray}\label{eq:two-point}
    \Gamma^{(2)}_{\eta_{(2)}\eta_{(2)}}(\vec{q},t)&\approx& \frac{1}{2E_{\eta_{(2)}}(\vec{q})V}|Z_{\eta_{(2)}}(\vec{q})|^2 e^{-E_{\eta_{(2)}}(\vec{q})t},\nonumber\\
    \Gamma^{(2)}_{J/\psi J/\psi,ii}(t)&\approx&\frac{1}{2m_{J/\psi}V}|Z_{J/\psi}|^2 e^{-m_{J/\psi}t}.
\end{eqnarray}
Note that $\Gamma^{(2)}_{\eta_{(2)}\eta_{(2)}}(\vec{q},t)$ includes the contributions from both connected and disconnected diagrams. We can extract the matrix elements $\langle \eta_{(2)}|j_\mathrm{em}^\mu|J/\psi\rangle$. For the kinetic configuration in Eq.~(\ref{eq:three-point}), the explicit expression of $Q^2$ is $Q^2=|\vec{q}|^2-\left[m_{J/\psi}-E_{\eta_{(2)}}(\vec{q})\right]^2$. Thus for a given $\vec{q}$ we can extract the form factor $M(Q^2)$ using Eqs.~(\ref{eq:spectral}), (\ref{eq:two-point}), and (\ref{eq:matrix}).

We observe the dominance of $J/\psi$ on $\Gamma_{\mu i}^{(3)} (\vec{q};t,t')$ when $t'>40$. For each $Q^2$, we fix $t'=40$ to get $M(Q^2,t)$ when $t>t'$. Fig.~\ref{fig:formfactor-t} shows the $t$ dependence of $M(Q^2,t)$ at several $Q^2$ close to $Q^2=0$. It is seen that a plateau region appears beyond $t-t'>10$ for each $Q^2$, where $M(Q^2)$ is obtained through a constant fit. The solid lines illustrate the central values and fitting time ranges, while the dashed lines indicate the jackknife errors. We also try to use a function $M(Q^2,t)=M(Q^2)+c(Q^2) e^{-\delta E (t-t')}$ to fit the data points at smaller $t-t'$ (shaded bands in Fig.~\ref{fig:formfactor-t}). The exponential term is introduced to account for the higher state contamination. The fitted $M(Q^2)$ in this way are consistent with those in the constant fit but have much larger errors. Therefore, we use the results from the constant fit for the values of $M(Q^2)$, which are listed in Table~\ref{tab:result}.

\begin{figure}[t!]
    \includegraphics[width=0.9\linewidth]{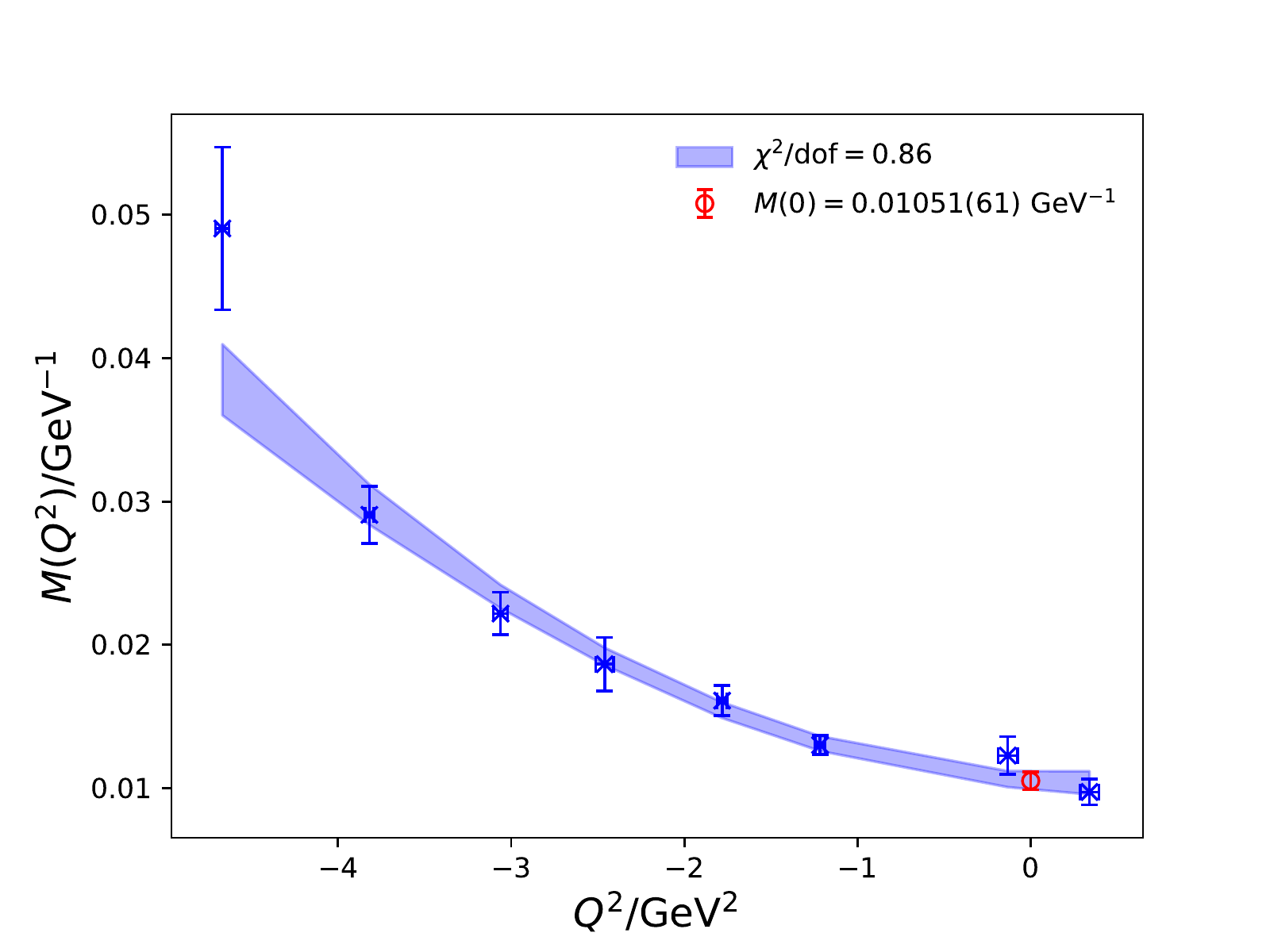}
    \caption{\label{fig:formfactor} From factor $M(Q^2)$ with respect to $Q^2$ in the physical unit. Data points indicate the numerical results. The bars present error values that are derived through jackknife resampling. The shaded curve illustrates the interpolation using the polynomial $M(Q^2)=M(0)+aQ^2+bQ^4$. The red circle with error bar exhibits the fitted form factor $M(0)=0.01051(61)$ GeV$^{-1}$.}
\end{figure}

In order to predict the partial decay width for the process $J/\psi\to \gamma \eta_{(2)}$ using Eq.~(\ref{eq:width}), the on shell form factor $M(Q^2=0)$ is required and can be obtained by the $Q^2$ interpolation of $M(Q^2)$ to $Q^2=0$. In practice, we perform a polynomial interpolation $M(Q^2)=M(0)+aQ^2+bQ^4$. The shaded curve in Fig.~\ref{fig:formfactor} exhibits that this function describes the dependence of $M(Q^2)$ on $Q^2$ very well in the available $Q^2$ range and gives the interpolated value $M(0)=0.01051(61)$ GeV$^{-1}$ (labeled as a red point in the figure). Plugging this value into Eq.~(\ref{eq:width}), the partial width and the branching fraction of the decay process $J/\psi\to \gamma \eta_{(2)}$ are predicted to be
\begin{eqnarray}\label{eq:branching}
    \Gamma(J/\psi\to \gamma \eta_{(2)})&=& 0.385(45)~~\mathrm{keV}\nonumber\\
    \mathrm{Br}[J/\psi\to \gamma \eta_{(2)}]&=& 4.16(49)\times 10^{-3},
\end{eqnarray}
where the branching fraction is deduced by using the $J/\psi$ total width $\Gamma=92.6(1.7)$ keV.

{\it Discussion.---}The branching fraction of the process $J/\psi\to \gamma \eta_{(2)}$ in Eq.~(\ref{eq:branching}) has been comparable with the experimental result $\mathrm{Br}[J/\psi\to \gamma\eta']=5.25(7)\times 10^{-3}$~\cite{Zyla:2020zbs}. However, a more appropriate comparison can be performed as follows. Firstly, $M(0)=0.01051(61)$ GeV$^{-1}$ for $\eta_{(2)}$ is close to $M(0)=0.0090(16)$ GeV$^{-1}$ for the pure gauge pseudoscalar glueball~\cite{Gui:2019dtm}; therefore, it does not show a clear $\mathcal{O}(\alpha_s^2)$ suppression expected for $q\bar{q}$ mesons. Secondly, it is observed in experiments that pseudoscalar mesons, such as $\eta'$, $\eta(1405/1475)$, $\eta(1760)$, $X(1835)$, and $\eta(2225)$, usually have large branching fractions~\cite{Zyla:2020zbs}, and their effective couplings in the $J/\psi$ radiative decay are close to each other in magnitude~\cite{Gui:2019dtm}. As such there may exist some general mechanisms behind these facts, among which the QCD $\mathrm{U_A(1)}$ anomaly can be most important since it enhances the gluon-pseudoscalar coupling nonperturbatively, as manifested by the anomalous axial current relation in the chiral limit
\begin{equation}\label{eq:anomaly}
    \partial_\mu j_5^\mu(x)=\sqrt{N_f} \frac{g^2}{32\pi}G_{\mu\nu}^a(x) \tilde{G}^{a,\mu\nu}(x)\equiv \sqrt{N_f} q(x),
\end{equation}
where $j_5^\mu=\frac{1}{\sqrt{N_f}}\sum\limits_{k=1}^{N_f} \bar{q}_k\gamma_5\gamma^\mu q_k$ is the flavor singlet axial vector current for $N_f$ flavor quarks, and $q(x)$ is the topological charge density. In the meantime, Eq.~(\ref{eq:anomaly}) also indicates that the anomalous gluon-pseudoscalar coupling is proportional to $\sqrt{N_f}$. Therefore, if $M(0)$ is dominated by the anomaly, we have the approximate relation $M(0,N_f=3)=\sqrt{3/2} M(0,N_f=2)$ and get an estimate $M(0)=0.01287(75)$ GeV$^{-1}$ for the physical $\textrm{SU(3)}$ case. On the other hand, for the physical $\textrm{SU(3)}$ flavor symmetry case, the physical $\eta$ and $\eta'$ are mass eigenstates and are admixtures of the flavor singlet $\eta_1$ and the flavor octet $\eta_8$, namely
\begin{eqnarray}
    |\eta\rangle&=&\cos \theta |\eta_8\rangle - \sin\theta |\eta_1\rangle,\nonumber\\
    |\eta'\rangle&=&\sin \theta |\eta_8\rangle + \cos\theta |\eta_1\rangle,
\end{eqnarray}
where $\theta$ is the mixing angle. Considering the mixing effects and using the physical $m_\eta=548$ MeV and $m_{\eta'}=958$ MeV, we can predict the branching fraction of $J/\psi\to \gamma \eta'$ as
\begin{eqnarray}\label{eq:lin}
    \mathrm{Br}[J/\psi\to \gamma \eta]&=& 1.15(14)\times 10^{-3}\nonumber\\
    \mathrm{Br}[J/\psi\to \gamma \eta']&=& 4.49(53)\times 10^{-3}
\end{eqnarray}
for $\theta_\mathrm{lin}=-24.5^\circ$ from the linear Gell-Mann-Okubo (GMO) mass relation~\cite{Zyla:2020zbs}, and
\begin{eqnarray}
    \mathrm{Br}[J/\psi\to \gamma \eta]&=& 0.256(30)\times 10^{-3}\nonumber\\
    \mathrm{Br}[J/\psi\to \gamma \eta']&=& 5.21(62)\times 10^{-3}
\end{eqnarray}
for $\theta_\mathrm{quad}=-11.3^\circ$ from the mass squared GMO relation~\cite{Zyla:2020zbs}. Obviously, the production rate of $\eta$ is very sensitive to $\theta$. With the consideration of the experimental branching fraction  $\mathrm{Br}[J/\psi\to \gamma \eta]=1.11(3)\times 10^{-3}$, the result from $\theta_\mathrm{quad}$ is too small, while from $\theta_\mathrm{lin}$ almost reproduces the experimental value within the error. This indicates that it is more proper to use $\theta_\mathrm{lin}$ here. We notice a recent sophisticated lattice study on $\eta$ and $\eta'$~\cite{Bali:2021qem} has calculated the matrix elements $a_{\eta(\eta')}$ of the topological charge density $q(x)$ between the vacuum and the $\eta(\eta')$ state, from which the mixing angle in the gluonic sector is derived to be $\theta_g=-\arctan \frac{a_\eta}{a_{\eta'}}\approx -24(4)^\circ$ at the energy scale $\mu=2$ GeV. If the $\mathrm{U_A(1)}$ anomaly dominates the decay process $J/\psi\to \gamma \eta(\eta')$, one should use $\theta_g$ to derive the branching fractions of $\eta$ and $\eta'$ from our result, which should be close to the values in Eq.~(\ref{eq:lin}) and agree well with experimental results. This manifests that the $\mathrm{U_A(1)}$ anomaly plays a crucial role in the $J/\psi$ radiative decay. The importance of the $\mathrm{U_A(1)}$ anomaly was also observed in the lattice study of the $D_s\to \eta'$ semileptonic decay~\cite{Bali:2014pva} where the contribution of disconnected quark diagrams is comparable to that of connected diagrams.

    {\it Summary.---}We have performed the first lattice study on the process of $J/\psi$ radiatively decaying to the isoscalar pseudoscalar $\eta_{(2)}$ on $N_f=2$ lattice QCD at $m_\pi\approx 350$ MeV. The involved light quark annihilation diagram is calculated by the distillation method. With a very large gauge ensemble consisting of about 7000 configurations, we obtain good signals for the desired three-point correlation functions with the insertion of the electromagnetic current. $m_{\eta_{(2)}}=717.7(8.3)$ MeV is measured by fitting the dispersion relation of $\eta_{(2)}$ on this lattice. Through the extracted form factor $M(0)=0.01051(61)$ GeV$^{-1}$, the partial decay width $\Gamma(J/\psi\to \gamma \eta_{(2)} )$ and the corresponding branching fraction are predicted to be 0.385(45) keV and $4.16(49)\times 10^{-3}$, respectively. By assuming that the $\mathrm{U_A(1)}$ anomaly is a dominance of the decay and considering the $\eta'-\eta$ mixing, our result provides the theoretical predictions for the production rate of the physical $\eta$ and $\eta'$ mesons in the $J/\psi$ radiative decay, which are in good agreement with the experimental values when the mixing angle is fixed at $\theta_\mathrm{lin}=-24.5^\circ$. In the present stage, we have only one lattice spacing, the uncontrolled systematic uncertainties, such as the $\mathrm{SU(2)}$ approximation, the chiral extrapolation and the continuum limit should be tackled in the future. Our result indicates the promising potential for lattice QCD to investigate light hadron productions in the $J/\psi$ radiative decay.

\vspace{2 cm}

This work is supported by the National Key Research and Development Program of China (No. 2020YFA0406400), the Strategic Priority Research Program of Chinese Academy of Sciences (No. XDB34030302), and the National Natural Science Foundation of China (NNSFC) under Grants No.11935017, No.12075253, No.12070131001 (CRC 110 by DFG and NNSFC), No.12075176, and No.12175063. The Chroma software system~\cite{Edwards:2004sx} and QUDA library~\cite{Clark:2009wm,Babich:2011np} are acknowledged. The computations were performed on the HPC clusters at the Institute of High Energy Physics (Beijing) and China Spallation Neutron Source (Dongguan), and ORISE Supercomputer.

\bibliography{ref}

\end{document}